\documentclass[aps,prl,twocolumn,groupedaddress]{revtex4}

\usepackage{subfigure}
\usepackage{graphicx}

\begin{document}

\title[ICP polishing of silicon for high quality optical resonators on a chip]{ICP polishing of silicon for high quality optical resonators on a chip}

\author{A. Laliotis$^1$\footnote{Present address: Laboratoire de Physique des Lasers, UMR 7538 du CNRS, Universite Paris-13, F-93430, Villetaneuse, France.}, M. Trupke$^1$\footnote{Present address: Atominstitut der Osterreichischen Universitaten, Stadionallee 2, 1020 Wien, Austria.}, J. P. Cotter$^1$, G. Lewis$^2$, M. Kraft$^2$ and E. A. Hinds$^1$}

\address{$^1$ The Centre for Cold Matter, The Blackett Laboratory, Imperial College London, SW7 2AZ, UK.}
\address{$^2$ School of Electronics and Computer Science, University of Southampton, Highfield, Southampton, SO17 1BJ, UK}
\email{j.cotter@imperial.ac.uk}
\email{ed.hinds@imperial.ac.uk}

\begin{abstract}
Miniature concave hollows, made by wet etching silicon through a circular mask, can be used as mirror substrates for building optical micro-cavities on a chip. In this paper we investigate how ICP polishing improves both shape and roughness of the mirror substrates. We characterise the evolution of the surfaces during the ICP polishing using white-light optical profilometry and atomic force microscopy. A surface roughness of {\boldmath 1}\,nm is reached, which reduces to {\boldmath 0.5}\,nm after coating with a high reflectivity dielectric. With such smooth mirrors, the optical cavity finesse is now limited by the shape of the underlying mirror.
\end{abstract}

\maketitle

\section{Introduction}
Microfabricated atom chips\cite{atomchipbook} that manipulate ultracold atoms are becoming increasingly important in a variety of applications, such as clocks\cite{treutlein04}, Bose-Einstein condensates\cite{hansel01}\cite{ott01}, matter wave interferometers\cite{schumm05}\cite{baumgartner10}, and quantum metrology\cite{riedel10}. These chips miniaturise cold atom experiments into small packages, integrating functions such as trapping\cite{lewis09,pollock09,pollock11}, guiding\cite{hinds99,dekker00} and detecting cold atoms\cite{eriksson05}. There is great interest now in integrating optical resonators into an atom chip in order to achieve strong atom-photon coupling for atom detection\cite{colombe07,goldwin11} and other applications in quantum information processing\cite{lepert11}. The desire for high quality optics integrated on chips has created the need for new miniature optical components and for fabrication techniques to shape their surfaces accurately and make them smooth.

\begin{figure}[h!]
\includegraphics[width=0.45\columnwidth]{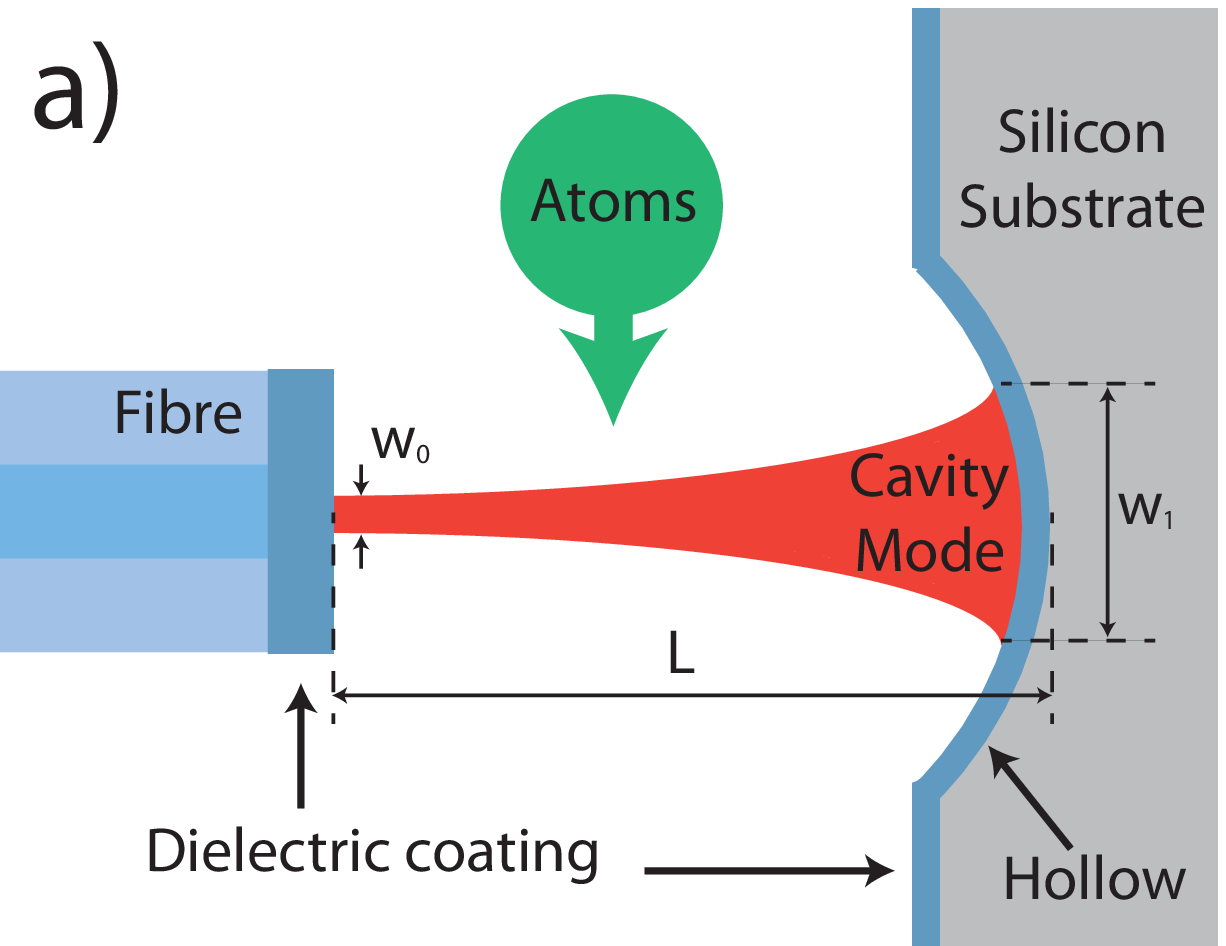}
\hspace{0.5cm}
\includegraphics[width=0.45\columnwidth]{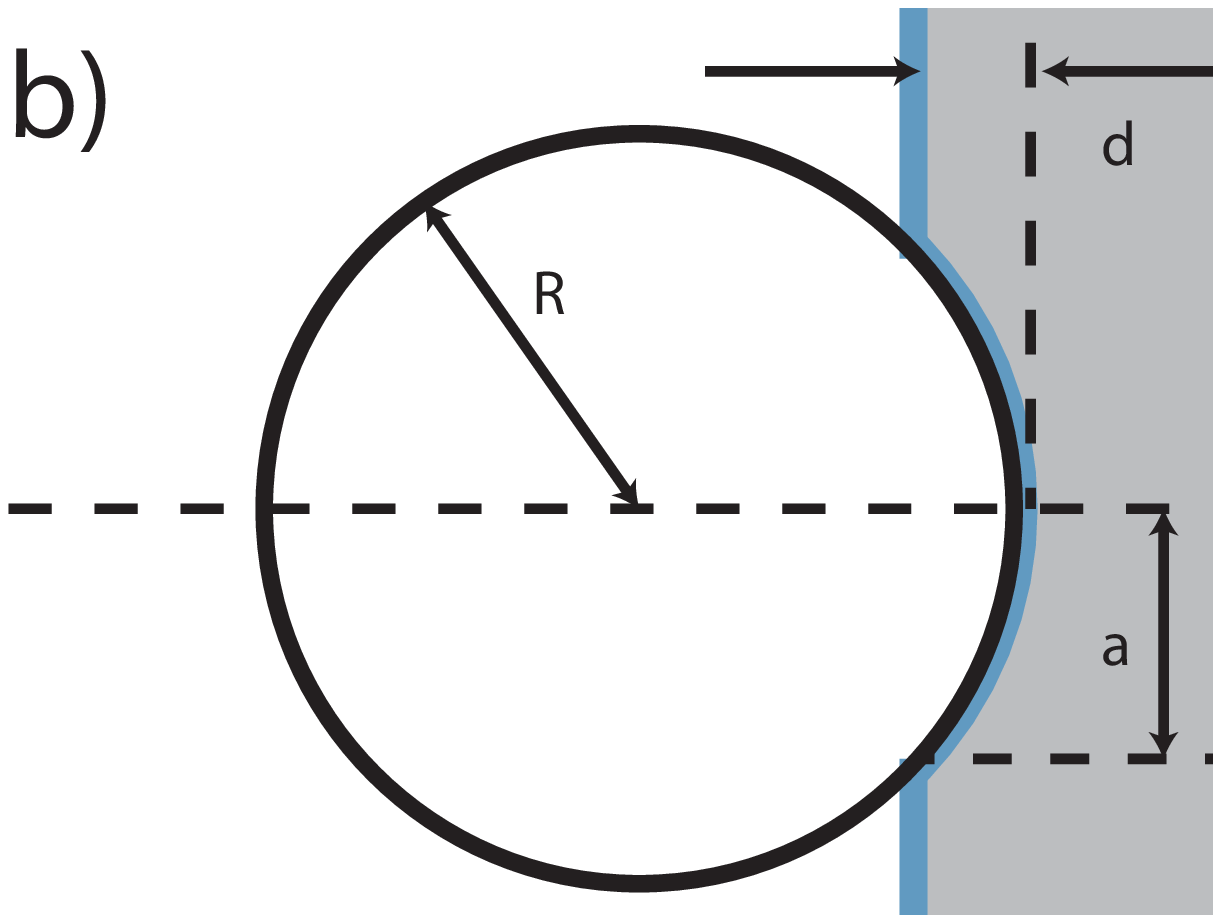}
\caption{a) Optical micro-cavity. The end of an optical fibre and a silicon hollow are coated with a dielectric multi-layer to create an optical resonator of length $L$, typically $100\,\mu$m. An atom passing through the the cavity is detected because it changes the reflected power in the fibre. b) Geometry of the hollows. Cavities are made from smooth hollows with radius of curvature $R$. Other relevant parameters are the depth, $d$, and the opening radius $a$.
\label{fig:experiment}}
\end{figure}
One such cavity consists of a concave mirror etched into a silicon substrate, together with a plane mirror on the end of a cleaved optical fibre, as illustrated in Fig. \ref{fig:experiment}a. Light leaves the fibre with a waist of $w_{0} \sim 5\,\mu$m. Because of diffraction this expands until it reaches the curved mirror when it has a spot size of $w_{1}$, typically between $5\,\mu$m and $10\,\mu$m. Both the fibre and the curved mirror are coated with multi-layer dielectric stacks to achieve high finesse. The fabrication and characterisation of this cavity are described in\cite{trupke05} where we find that the cavity performance is limited by surface roughness. Despite this, such cavities with a finesse of a few hundred have already been used to detect atoms on a chip with high fidelity and fast response time\cite{trupke05,trupke07}.

It is known that inductively coupled plasma (ICP) etching with precise control over the etching parameters can polish surfaces\cite{larsen05,larsen06}. In this paper we investigate the effects of unmasked ICP polishing on cavity mirrors that have been created by wet etching in a solution of hydrofluoric, nitric and acetic acids (HNA). Detailed measurements of the shape and roughness are made at various times throughout the ICP polishing. We characterise the shape by the radius $a$ of the aperture, the depth $d$, and the radius of curvature $R$, shown in Fig. 1b. We find  an improvement in both the profile and the roughness of the silicon surface, making it a more suitable mirror substrate for optical micro-cavities of high finesse.

This paper is organised as follows: Section II describes the process used to fabricate the hollows and compares hollows of various sizes. Section III describes the ICP polishing process and its effect on the shape of the hollows. In section IV we investigate how the roughness of the silicon surface evolves during ICP polishing. We conclude with a discussion of the optical mirror quality and its application to optical micro-cavities.

\section{Wet-etching the hollows}

We started with 20 $\langle100\rangle$-oriented, P-type silicon wafers of $600\,\mu$m thickness on which a $60\,$nm layer of silicon nitride was deposited by low-pressure chemical vapour deposition (LPCVD). Each wafer was then primed using hexamethyldisilazane (HMDS) vapour and coated with a $1.3\,\mu$m thick HPR$-504$ photoresist, which we lithography patterned with circles of radii $10,20,30$ and $40\,\mu$m. The silicon nitride layer was then opened by a 20-second isotropic reactive ion etch (RIE) using $515\,$W of RF power, with a gas flow of $5\,$sccm CHF$_{3}$, $25\,$sccm CF$_{4}$ and $60\,$sccm Ar at a gas pressure of 40\,mTorr. The etch rates for silicon and silicon nitride are very similar, so this step requires care to avoid over-etching into the silicon wafer. The hollows were then fabricated by wet etching all the wafers for $2\,$minutes at room temperature under constant agitation, using an HNA solution of (HF:HNO$_{3}$:CH$_{3}$CO$_{2}$H) in the ratio (30:43:27) with concentrations of ($49$:$70$:$99.5$)\%wt. Although the HNA etch rate and the resulting hollow profiles are highly dependant on the composition and agitation of the solution\cite{kovacs98} this recipe gave repeatable surface profiles with approximately $6\,$nm rms roughness in the hollows\cite{trupke05}.

\begin{figure}[t]
\centering
\includegraphics[width=0.9\columnwidth]{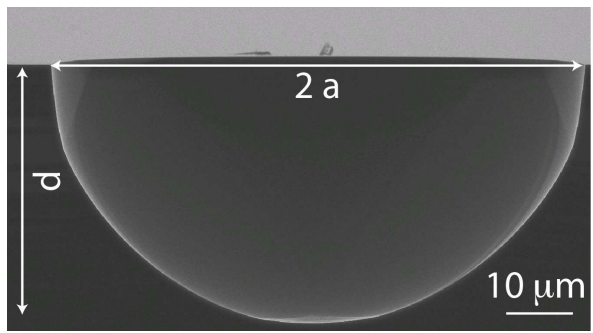}
\caption{SEM image showing the cross section of a hollow, wet-etched through a 10\,$\mu$m-radius opening (prior to ICP polishing). The bottom of the hollow is nearly spherical.}
\label{fig:cavSEM}
\end{figure}

\begin{figure}[b]
\centering
\includegraphics[width=0.9\columnwidth]{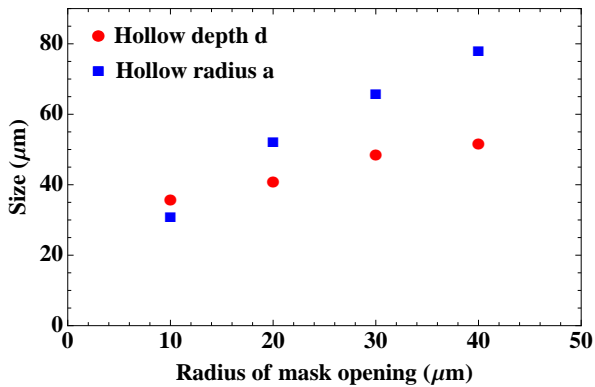}
\caption{Radius $a$ (blue squares) and depth $d$ (red circles) of the hollows as a function of original mask opening radius.}
\label{fig:cavSize}
\end{figure}
Fig. \ref{fig:cavSEM} is a scanning electron microscope image showing the cross section of the hollow made using a $10\,\mu$m-radius opening in the nitride mask. It has an aperture of $a\simeq 30\,\mu$m and a depth $d\simeq a$. The surface is roughly spherical at the bottom of the hollow. In order to see how these parameters vary with the size of the mask opening, all the hollows were measured using a stylus profilometer, and their dimensions are plotted in Fig.\,\ref{fig:cavSize} versus mask opening. We see that the depth does not increase as rapidly as the radius, making the larger hollows less than a hemisphere.
\begin{figure}[h!]
\centering
\subfigure{\includegraphics[width=0.9\columnwidth]{Fig4a.eps2}}
\vspace{0cm}
\subfigure{\includegraphics[width=0.9\columnwidth]{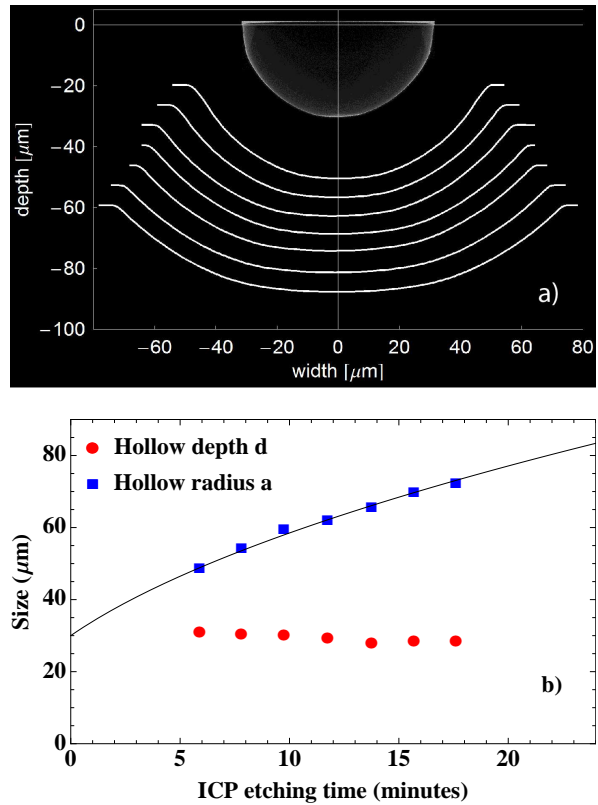}}
\caption{Evolution under ICP polishing of a hollow made by wet-etching through a $10\,\mu$m-radius nitride opening. a) Top profile: SEM image reproduced from Fig.\,\ref{fig:cavSEM}. Lower profiles: measured by stylus after ICP etching for $6, 8, 10, 12, 14, 16$, and 18 minutes. b) Radius $a$ (blue squares) and depth $d$ (red circles) of the hollows, plotted as a function of ICP polishing time. We see that the ICP etch does not change the depth of the hollow significantly but does increase the radius $a$. The line is a fit to a simple theory given in the text.
\label{fig:Stylus}}
\end{figure}

\section{Effect of ICP polishing on the shape}

The reactive ion etch, previously used to open holes in the nitride mask, was used again to remove the rest of the nitride. Some silicon is inevitably removed at the end of this step, slightly roughening the silicon surface, so care was taken to use the minimum etching time required to remove the nitride. This was preferable to a conventional orthophosphoric acid dip, which caused visible degradation of the silicon surface. Next, the wafers were ICP polished, in an STS Advanced Silicon Etcher using a chuck temperature of $20^{\circ}$C, $200\,$sccm SF$_{6}$ at a pressure of $10\,$mTorr, $16\,$W platen power and a coil power of $3000\,$W \cite{larsen05}. Each  wafer was polished for a different length of time, spanning the range $0-38\,$minutes in steps of 2\,minutes.

The profiles of the hollows were measured again after polishing. Fig. \ref{fig:Stylus}a shows the evolution of the hollow formed initially through a $10\,\mu$m-radius mask opening. The top profile is the initial SEM section already shown in Fig.\,\ref{fig:cavSEM}, while the lower images are stylus profiles taken after $6, 8, 10, 12, 14, 16$, and 18\,minutes of ICP polishing.  We were unable to obtain stylus profiles for shorter times due to the steep vertical walls. These measurements gave the values of $a$ and $d$ that are plotted versus ICP polishing time in Fig. \ref{fig:Stylus}b. We see that $d$ remains essentially constant, indicating that the etch rate is the same at the bottom of the hollow as it is on the flat surface of the wafer. Approximating the surface of the hollow as part of a sphere, the aperture radius at time $t$ is $a(t)=\sqrt{2 R(t) d-d^{2}}$, where $R(t)$ is the radius of curvature of the sphere. For an isotropic etch at rate $\Gamma$ we expect $R(t)=R(0)+\Gamma t$ \cite{larsen05}. The line in Fig. \ref{fig:Stylus}b is a fit to this simple model, showing good agreement with the data and yielding an ICP etch rate of $\Gamma=4.2(1)\,\mu$m/min.

\begin{figure}[t]
\centering
\includegraphics[width=0.9\columnwidth]{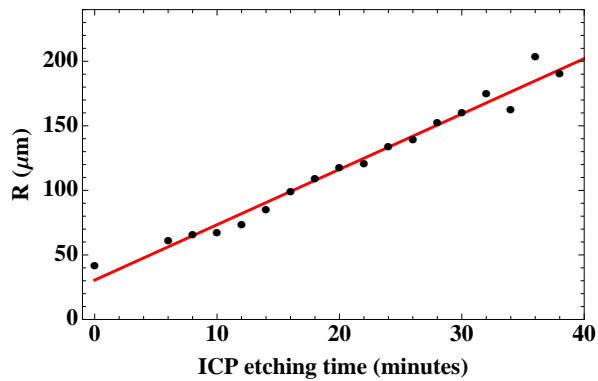}
\caption{Radius of curvature, $R$, as a function of ICP etch duration. There is a linear increase of $4.3(2)\,\mu$m/min in the radius of curvature with increasing ICP etching.
\label{fig:RadiusFItPlot}}
\end{figure}

A Zygo white-light interferometer allowed us to measure the hollows with higher resolution to check their suitability as mirror substrates for a high-finesse optical cavity. This profilometer gives a two dimensional map of height, on a grid of 150\,nm square pixels. We concentrate on a circular region of diameter $11.6\,\mu$m at the bottom of the hollows, a little larger than the spot size $w_{1}$ of the resonant microcavity mode, and small enough to be seen clearly, even in the smallest of the hollows. We use these images to determine $R(t)$ as a function of ICP polishing time, as shown in Fig.\,\ref{fig:RadiusFItPlot}, together with a straight line fit. The radius does indeed grow linearly, with the best fit to the growth rate being $\Gamma = 4.3(2)\,\mu$m/min, in agreement with the rate deduced above from the aperture size.

\begin{figure}[h!]
\centering
\includegraphics[width = 0.9\columnwidth]{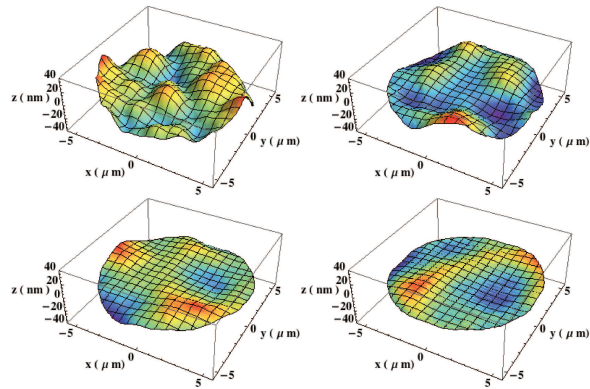}
\caption{Deviation of hollow surfaces from spherical, measured over a circular aperture of $11.6\,\mu$m diameter at the bottom of the hollows. These graphs show the residual  surfaces from a fit to a spherical section after ICP etching for $0$, $12$, $24$ and $38\,$minutes. The residuals initially span a range $\pm15\,$nm and are reduced after 38 mins of etching to $\pm3\,$nm, while the correlation length increases from $\sim 2\,\mu$m to $\sim 4\,\mu$m.
\label{fig:residuals}}
\end{figure}

The hollows deviate from being spherical, as plotted in Fig. \ref{fig:residuals}. Here we show four representative samples, which have been etched for $0$, $12$, $24$ and $38\,$ minutes. Initially the residuals vary typically over the range $\pm 15\,$nm, with a correlation length of $\sim 2\,\mu$m. After 38 minutes of ICP etching, the residuals are typically within the range  $\pm3\,$nm while the correlation length has doubled. This amounts to an order of magnitude reduction in the noise power of the residuals and almost two orders of magnitude in the angular noise power. Mirrors with this level of sphericity have been used in our laboratory to make cavities with finesse exceeding 6,000.

\section{Measurements of roughness}
\begin{figure*}[t!]
\centering
\includegraphics[width=1.95\columnwidth]{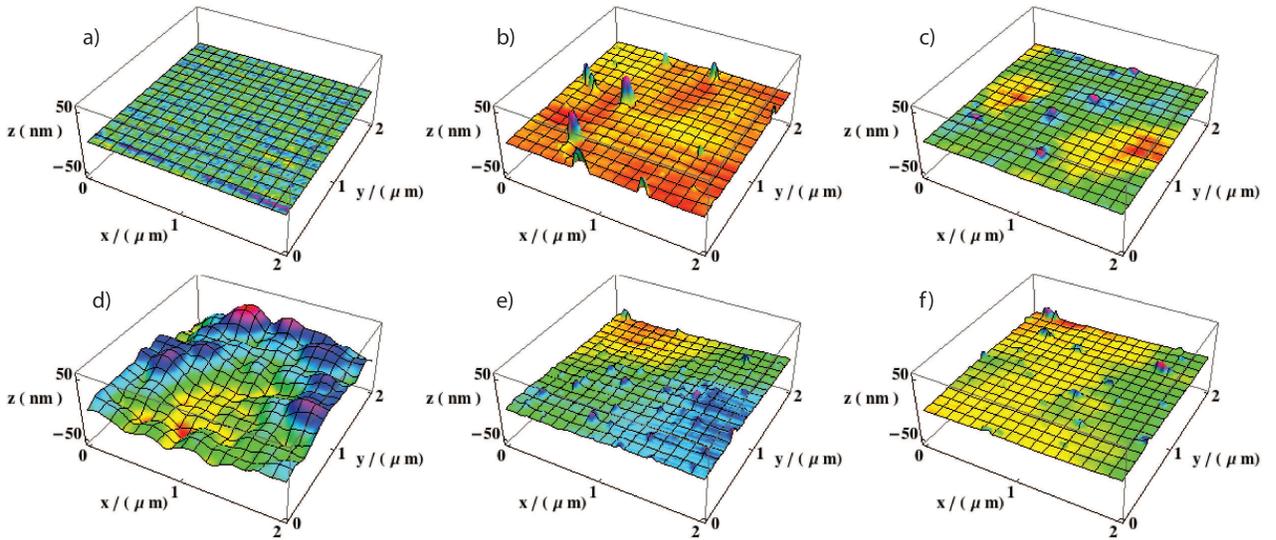}
\caption{AFM scans depicting the evolution of the surface under ICP polishing. a) - c) Flat part of the wafer. a) Ihe initial roughness is negligible (rms = 0.2\,nm). b) At $10$\,minutes, unknown contaminants are clearly seen on the surface. c) After $38\,$minutes they are almost entirely removed. d) - f) Deviation of the hollows from a sphere. d) There are large initial deviations from spherical. e) At $18\,$minutes the shape is improved, but contaminants can be seen, as on the flat. f) After $38\,$minutes these are largely removed. The polished surface has rms roughness $\leq1\,$nm.
\label{fig:afm}}
\end{figure*}
\begin{figure}[h!]
\centering
\subfigure{\includegraphics[width=0.9\columnwidth]{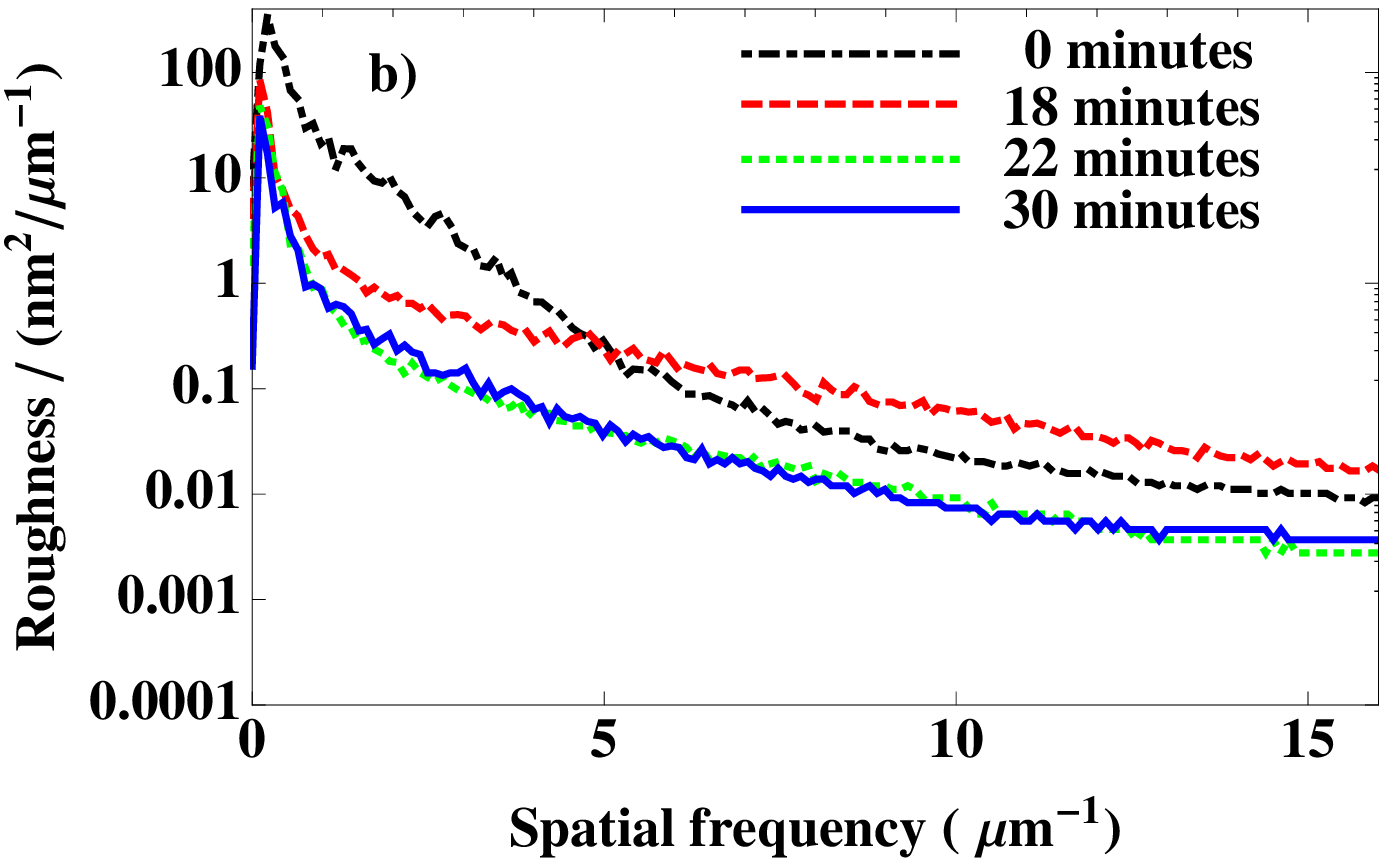}}
\subfigure{\includegraphics[width=0.9\columnwidth]{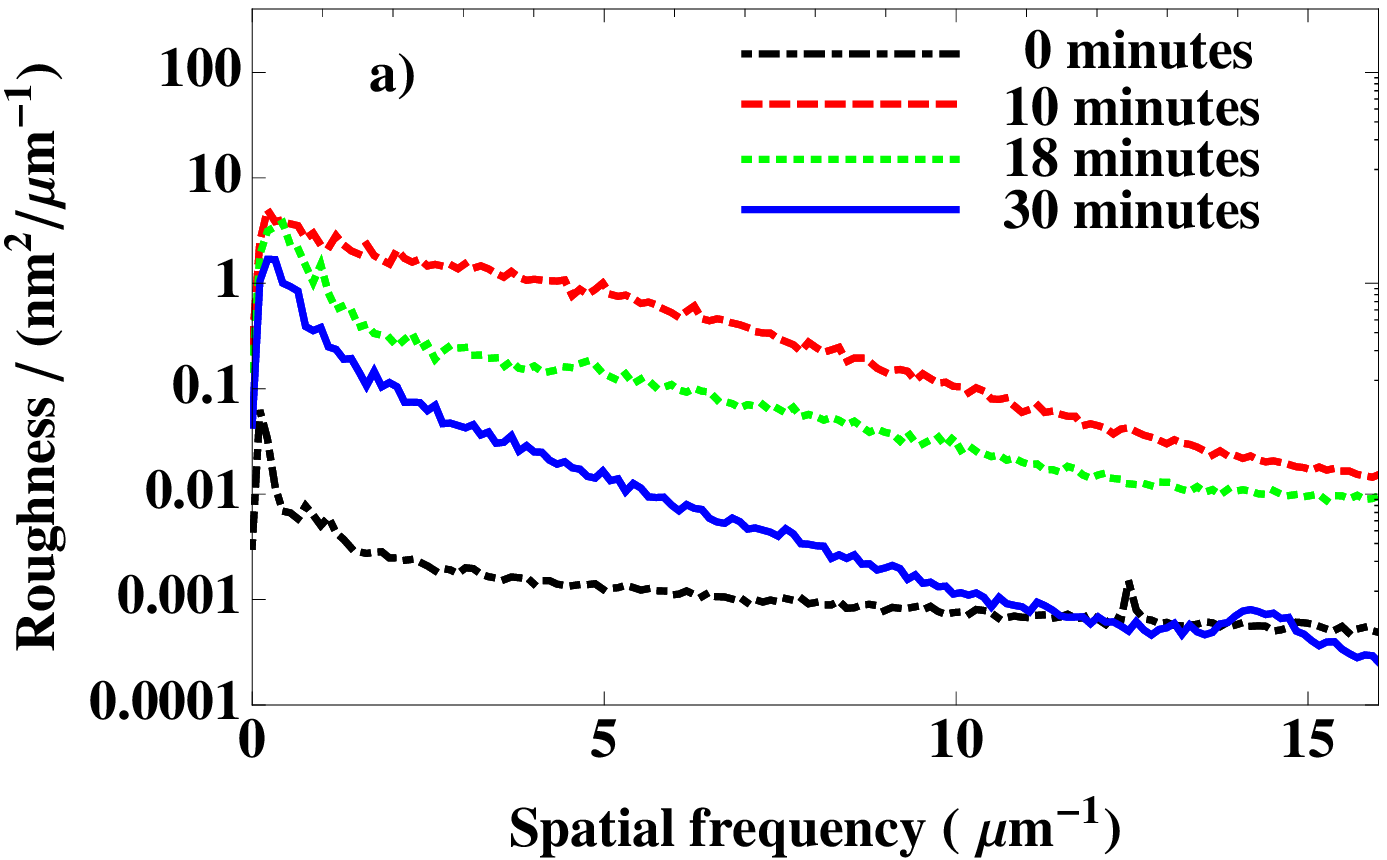}}
\caption{a) Roughness power spectrum of the flat part of the wafer after $0$ (dash-dotted, black), $10$ (dashed, green), $18$ (dotted, red) and $38\,$minutes (solid, blue) of ICP etching. b) Roughness power spectrum of the curved hollows after $0$ (dash-dotted, black), $18$ (dotted, red) and $38\,$minutes (solid, blue) of ICP etching.
\label{fig:Fafm}}
\end{figure}

In order to complete our picture of the polishing, we used an atomic force microscope (AFM) to see the roughness with much higher transverse resolution over a $10\,\mu$m$\times10\,\mu$m square. Figures\,\ref{fig:afm}(a - c) show scans on the flat part of the wafer, zoomed in to a $2\,\mu$m$\times2\,\mu$m region. Before ICP polishing, this part the wafer has no appreciable roughness, the standard deviation of the height being only $0.2\,$nm over the whole inverse wavelength band of $(0.1-25)\,\mu\mbox{m}^{-1}$. During the first $10-15$ minutes of polishing however, the standard deviation grows to $3.5\,$nm due to the appearance of particles - typically $(20-40)\,$nm in diameter with a density of order 2 per $\mu\mbox{m}^2$ - whose origin is unknown. After that, the contaminant particles are gradually removed and the standard deviation decreases again. At 38 minutes, these particle are still present, but their height is reduced to a few nm, giving an rms variation of $1.1\,$nm. Discounting them, the roughness of the underlying surface is approximately 0.5\,nm. Figure\,\ref{fig:Fafm}(a) plots the average power spectrum of height noise measured along the $x$ direction (the $y$ direction gives the same result). Here the arrival of the localised contaminant particles shows up as a broadband increase in the noise power density, followed by a gradual return to a lower noise, particularly at high spatial frequency. The roughening of the underlying surface prevents the low-frequency noise power density from returning to its  original value after 38 minutes.

Figure\,\ref{fig:afm}(d -f) shows a similar set of scans, this time at the base of the hollow. In (d), the initial unpolished profile deviates from spherical over the range $\pm15$\,nm, as already seen with the Zygo profilometer. This image could only be made after cleaving the waver as the opening is otherwise too small to accommodate the AFM cantilever. The spectral density of the noise is plotted in Fig.\,\ref{fig:Fafm}(b). We see that the roughness is predominantly due to Fourier components at long wavelength ($\gtrsim 1\,\mu$m) associated with the shape of the hollow, rather than the microscopic roughness of the surface. After 18 minutes of polishing, the hollow is large enough to measure without cleaving. The scan of this hollow shows that the large scale structure is becoming smoother and, in addition, we see the same covering of contaminant particles found on the flat surface. In the Fourier spectrum this appears as a decrease in the low-frequency noise power and an increase at high frequency, i.e. above $4\,\mu\mbox{m}^{-1}$. After 38 minutes, the particle contamination has been largely cleaned away, as on the flat surface, and the deviations of the surface from spherical are reduced to the typical range $\pm 2\,$nm, as also seen with the Zygo profilometer. The low frequency noise is the polished relic of the much larger shape imperfection seen initially. In the spatial frequency band above an inverse wavelength of $1\,\mu\mbox{m}^{-1}$, the variance of the surface height is less than $1\,\mbox{nm}^2$.

\section{Conclusions}
\label{sec:summary}
We have found that the hollow prepared by wet etching has an undulating surface that deviates from spherical by typically $\pm15\,$nm over distances of order $2\,\mu$m. The ICP polishing reduces these low-frequency deviations dramatically. As far as an optical cavity is concerned, this part of the spectrum describes the deviation of the surface from spherical. It may be regarded as noise in the shape of the mirror substrate, producing geometrical aberrations of the cavity that were also seen using the white light interferometer. In the frequency band above $1\,\mu\mbox{m}^{-1}$, there is an initial increase of noise due to the deposition of contaminant particles on the surface. After 38 minutes, these particles are largely polished away and the variance of the surface is decreased to $\le 1\,\mbox{nm}^2$. These higher frequency Fourier components are mostly missed by the optical profilometer because they are not resolved by the light and merely serve to reduce the reflectivity of the surface from it's ideal value $\rho_{0}$. The Debye-Waller formula\cite{trupke05}, $\rho =\rho_{0} \exp{[- (4 \pi \sigma / \lambda)^{2}]}$, provides a simple estimate of the reduced reflectivity, where $\sigma$ is the surface roughness and $\lambda$ is the wavelength of the light. A typical application of the cavity is in cavity QED\cite{trupke07}. For example, using the D line of Cs atoms at $852\,$nm, a roughness of 1\,nm permits a reflectivity of 99.98\%, corresponding to a cavity finesse ($\pi\sqrt{\rho}/(1-\rho)$ with matched mirrors) exceeding 14,000 for $\rho_{0}=1$. However, the reflectivity  $\rho_{0}$ is low in the visible and near-infrared because of absorption by the silicon, so the hollow has to be coated by a dielectric Bragg stack several microns thick, depending on the required reflectivity. The surface roughness of this coating is below $0.5\,$nm which can support a reflectivity of 99.995 \% or a finesse exceeding 60,000.

In search of the highest finesse, Biedermann \textit{et al.}\cite{biedermann10} added a final silicon polishing stage, in which a thick oxide was twice grown on the surface and removed by HF. This left the silicon surface with a roughness of $\sigma=0.22\,$nm at spatial frequencies above $1\,\mu\mbox{m}^{-1}$. The roughness of the Bragg stack deposited on top of this was $0.26\,$nm, slightly better than the roughness of the optical coating in our laboratory. The maximum finesse they achieved was 64,000, impressively high but much less than the Debye-Waller maximum of 260,000.

We conclude that our technique for polishing silicon provide a smooth substrate, suitable for building micro-mirrors of high reflectivity corresponding to a finesse in excess of 60,000. Even smoother surfaces can be achieved, but at this point, the limiting imperfection is the shape of the surface, rather than its roughness. This might be improved by even longer ICP etching time, but a better approach is probably to replace the initial wet etch by an ICP etch.

\section*{Acknowledgements}
This work was supported by the CEC Seventh Framework projects 247687 (AQUTE) and 221889 (HIP), the UK EPSRC and The Royal Society. We are indebted to Teng Zhao for help with the AFM imaging.

\section*{References}
\bibliographystyle{unsrt}

\end{document}